\documentclass[aps, prb, twocolumn, amsmath, amssymb, reprint, groupedaddress]{revtex4-1}
\usepackage{graphicx, hyperref, color}

\begin{document}

\title{Topological Properties of a Coupled Spin-Photon System Induced by Damping}

\author{Michael Harder}
\email{michael.harder@umanitoba.ca}

\author{Lihui Bai}

\author{Paul Hyde}

\author{Can-Ming Hu}

\affiliation{Department of Physics and Astronomy, University
of Manitoba, Winnipeg, Canada R3T 2N2}
 
\date{\today}

\begin{abstract}
We experimentally examine the topological nature of a strongly coupled spin-photon system induced by damping.  The presence of both spin and photonic losses results in a non-Hermitian system with a variety of exotic phenomena dictated by the topological structure of the eigenvalue spectra and the presence of an exceptional point (EP), where the coupled spin-photon eigenvectors coalesce.  By controlling both the spin resonance frequency and the spin-photon coupling strength we observe a resonance crossing for cooperativities above one, suggesting that the boundary between weak and strong coupling should be based on the EP location rather than the cooperativity.  Furthermore we observe dynamic mode switching when encircling the EP and identify the potential to engineer the topological structure of coupled spin-photon systems with additional modes.  Our work therefore further highlights the role of damping within the strong coupling regime, and demonstrates the potential and great flexibility of spin-photon systems for studies of non-Hermitian physics.   
\end{abstract}

\maketitle
\section{Introduction} \label{sec:introduction}

Strong coupling between microwave fields and spin excitations in ferrimagnetic materials has received significant attention in recent years [\!\!\!\citenum{Soykal2010, Huebl2013a, Zhang2014, Tabuchi2014a, Bai2015, Zhang2015e, Haigh2015a, Cao2014, Yao2015, Bourhill2015a}].  Such systems offer a combination of long coherence times, due to the low damping of certain ferrimagnetic materials, and the realization of large coupling strengths, which are enhanced by the ferrimagnetic ordering [\!\!\citenum{Soykal2010a, Cao2014}], making them exciting platforms for both quantum information [\!\!\citenum{Huebl2013a, Zhang2015g, Tabuchi2015b}] and spintronic [\!\!\citenum{Bai2015, Hu2015, Bai2017}] applications.  Indeed the recent proposal of a magnon dark mode memory architecture [\!\!\citenum{Zhang2015g}], demonstration of magnon-qubit coupling [\onlinecite{Tabuchi2015b}], the development of cavity optomagnonics [\onlinecite{Haigh2015b, Zhang2015b, Osada2015, Bourhill2015, Liu2016a, Hisatomi2016}] and the demonstration of non-local spin current control [\!\!\citenum{Bai2017}] exemplify both the intriguing physics and application potential of coupled spin-photon systems.  The key to the majority of these recent studies has been the ability to enter the so-called strong coupling regime, where a large spin-photon coupling strength ($\kappa$), compared to smaller spin ($\alpha$) and photonic ($\beta$) losses, leads to exceptionally large cooperativities, $C = \kappa^2/\alpha\beta >1$.  In this regime the coupling strength is typically characterized by an anti crossing in the frequency dispersion [\!\!\citenum{Harder2016b}] which can be well described by a two-level model [\!\!\citenum{Huebl2013a, Zhang2014, Cao2014, Bai2015, Yao2015, Harder2016b}], which is equivalent to a set of coupled oscillators [\!\!\citenum{Harder2016b}].  However interestingly, in the strong coupling regime this dispersion appears to be insensitive to the damping properties, despite the fact that damping plays such an important role in the very realization of strong coupling.

On the other hand, the presence of damping generally gives rise to non-Hermitian Hamiltonians and damping is therefore well known to play a fundamental role in the physics of two-level systems [\!\!\citenum{Heiss1999, Heiss2000, Bender1998, Bender2007}].  In particular the presence of damping may lead to a dispersion crossing even in the presence of coupling [\!\!\citenum{Heiss1999, Heiss2000, Philipp2000}].  Such crossing phenomena is tied to the intricate topology of the eigenvalue spectrum in non-Hermitian systems, and is controlled by the presence of an exceptional point (EP), where the eigenvectors of the non-Hermitian system coalesce [\!\!\citenum{Heiss1999, Heiss2000, Dembowski2001}].  Beyond controlling the crossing/anticrossing behaviour, the presence of an EP also leads to a variety of other exotic phenomena such as anomalous geometric phases [\!\!\citenum{Heiss1999, Heiss2000, Dembowski2001}], controllable coherence in lasing systems [\!\!\citenum{Liertzer2012}], non-reciprocal energy transfer [\!\!\citenum{Xu2016}] and the breakdown of adiabatic evolution in waveguides [\!\!\citenum{Graefe2013, Doppler2016}].  Yet despite the wealth of interesting phenomena, this exciting frontier of non-Hermitian physics has yet to be explored within the context of coupled spin-photon systems.

In this work we explore the role of damping in a strongly coupled spin-photon system and its relationship to the topological structure.  We find that the dispersion gap does display a damping dependence, which becomes important even within the strong coupling regime, resulting in the potential for a dispersion crossing when $C > 1$.  Such a crossing is experimentally observed when $C = 1.3$, suggesting that the location of the EP, rather than the cooperativity should be used to define the strong/weak transition.  Furthermore we demonstrate that the topological eigenvalue structure will lead to mode switching when the EP is encircled and propose a method to add exceptional points, thus tuning the topological structure, using additional cavity or spin resonance modes.  Therefore we find that the presence of an EP should play an important role in studies of spin-photon coupling near the strong/weak transition and that the versatility of such a system, with in-situ tuneable cavity damping, resonance frequencies, and coupling strength, as well as the potential to add both cavity and spin resonance modes, provides an intriguing playground for the exploration of non-Hermitian physics.  

We begin in Sec. \ref{sec:topological} by describing the basic topological structure of the spin-photon system, emphasizing the important role of damping and the non-Hermitian nature of our system, and present the experimental realization of a dispersion crossing within the strong coupling regime.  In Sec. \ref{sec:phase} we demonstrate the mode switching which occurs when the EP is encircled.  Finally in Sec. \ref{sec:manipulate} we demonstrate the versatility of the spin-photon system in engineering the topological structure by adding a second cavity mode, before concluding with Sec. \ref{sec:conclusion}.

\section{Topological Nature of The Coupled Spin Photon System} \label{sec:topological}

\begin{figure}[t!]
 \includegraphics[width = 8 cm]{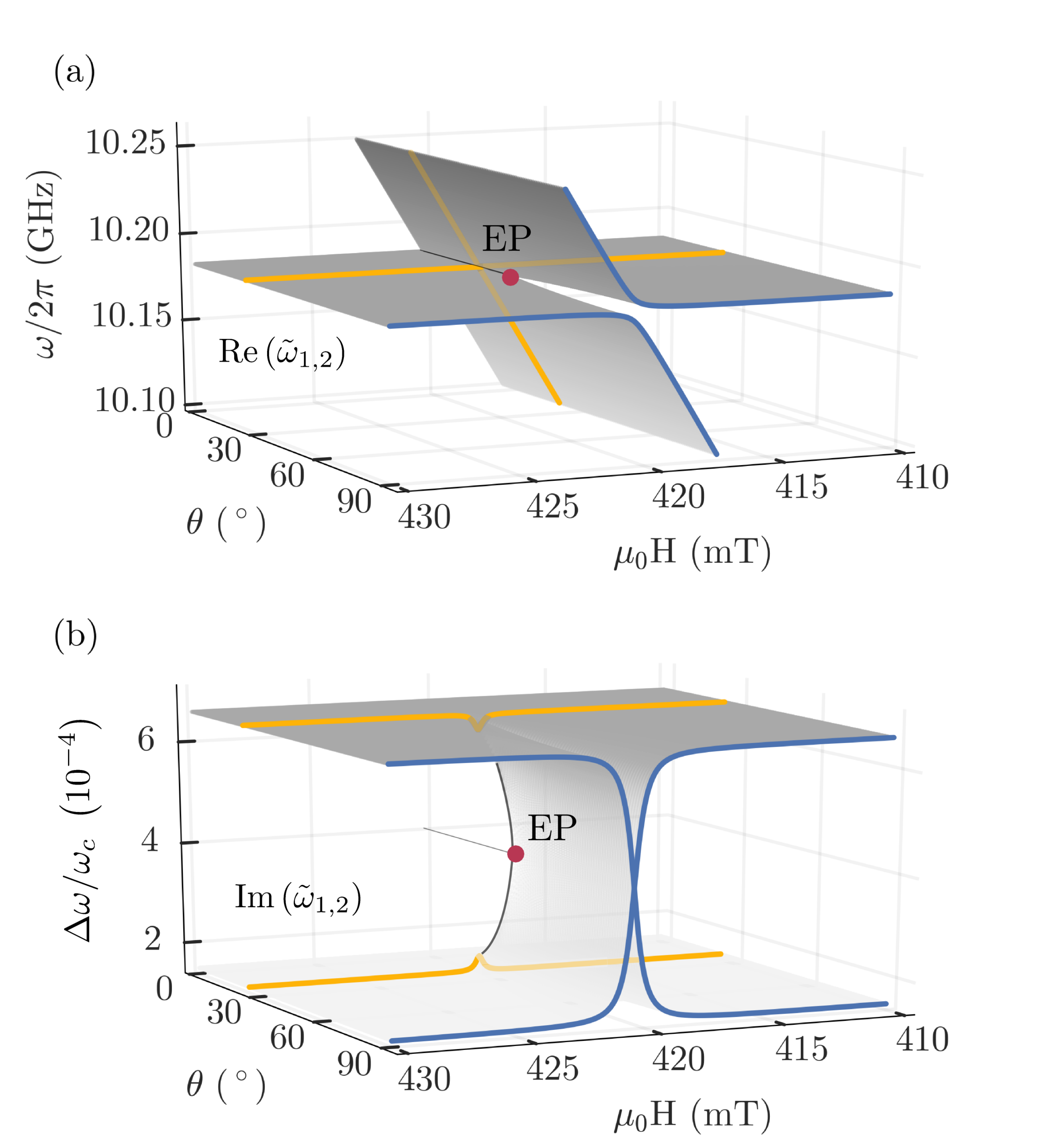}
\caption{The square root eigenvalue topology of the spin-photon system.  (a) The dispersion (b) and line width are calculated from $\tilde{\omega}_{1,2}$ according to Eq. \eqref{eq:eigen} using our experimental parameters.  Blue (dark) curves at $\theta > \theta_\text{EP}$ show the characteristic strong coupling behaviour; a dispersion anti crossing and line width crossing.  However when $\theta < \theta_\text{EP}$ the yellow (light) curves display a dispersion crossing and line width anti crossing, even though $C > 1$.  The red circle indicates the exceptional point.}
\label{fig1}
\end{figure}

The key physics of a coupled spin-photon system, comprised of a ferrimagnetic material and a microwave cavity/resonator, can be accurately modelled by the non-Hermitian two-level Hamiltonian [\!\!\citenum{Cao2014, Bai2015, Harder2016b}]
\begin{equation}
H = \left(\begin{array}{cc}
\omega_c - i \beta \omega_c & \kappa \omega_c \\
\kappa \omega_c & \omega_r - i \alpha \omega_c
\end{array} \right). \label{eq:hamiltonian}
\end{equation}
Here $\omega_c$ and $\omega_r$ are, respectively, the cavity and ferromagnetic resonance (FMR) frequencies and we have normalized the dimensionless coupling strength $\kappa$ and the cavity $(\beta)$ and FMR $(\alpha)$ damping parameters using $\omega_c$.  The complex eigenvalues, $\tilde{\omega}_{1,2}$, which determine the topological structure of the system, can be found by diagonalizing the two-level Hamiltonian,
\begin{equation}
\tilde{\omega}_{1,2} = \frac{1}{2} \left[\tilde{\omega}_c + \tilde{\omega}_r \pm \sqrt{\left(\tilde{\omega}_c-\tilde{\omega}_r\right)^2 + 4 \kappa^2 \omega_c^2}\right],\label{eq:eigen}
\end{equation}
where $\tilde{\omega}_c = \omega_c - i \beta \omega_c$ and $\tilde{\omega}_r = \omega_r- i \alpha \omega_c$ are the complex eigenvalues in the $\kappa \to 0$ limit.
At the coupling point, $H_c$, where $\omega_c = \omega_r (H_c)$, these eigenvalues produce a well known dispersion gap for very large coupling strengths, as illustrated by the blue (dark) curve in Fig. \ref{fig1} (a).  

To examine the damping dependence of the coupling gap we write $\tilde{\omega}_{1,2} = \omega_{1,2 } - i \Gamma_{1,2} \omega_c$ where $\omega_{1,2}$ is the real part of the eigenvalue which determines the dispersion relation (experimentally $\omega_{1,2} = \omega$ is the frequency), and $\Gamma_{1,2}$ is the imaginary part which characterizes the spin-photon lifetime (experimentally $\Gamma_{1,2} = \Delta \omega$ is the half width at half maximum).  From Eq. \eqref{eq:eigen} the difference between the two eigenmodes is,  
\begin{equation}
\left(\tilde{\omega}_1 - \tilde{\omega}_2\right)^2 = \left(\tilde{\omega}_c - \tilde{\omega}_r\right)^2 + 4 \kappa^2 \omega_c^2 \label{eq:cond1}.
\end{equation}
Taking the real and imaginary parts of Eq. \eqref{eq:cond1} yields the following two conditions which must be satisfied at the coupling point,
\begin{subequations}
\begin{gather}
\left(\omega_1 - \omega_2\right) \left(\Gamma_1 - \Gamma_2\right) = 0\label{eq:cond2} \\
\left[\frac{\omega_1-\omega_2}{\omega_c}\right]^2= \left(\Gamma_1 - \Gamma_2\right)^2 - \left(\beta - \alpha\right)^2 + 4 \kappa^2. \label{eq:cond3}
\end{gather}
\end{subequations}
Eq. \eqref{eq:cond2} will be satisfied if: 
\begin{align*}
&\text{(i)}~ \Gamma_1 = \Gamma_2, \\
&\text{(ii)} ~\omega_1 = \omega_2,~ \text{or}, \\
&\text{(iii)} ~\omega_1 = \omega_2~ \text{and} ~\Gamma_1 = \Gamma_2
\end{align*}
which corresponds to the following physical situations:
\begin{align*}
&\text{(i) Resonance anti crossing, line width crossing},  \\
&\text{(ii) Resonance crossing, line width anti crossing}, \\ 
&\text{(iii) Resonance crossing, line width crossing}.
\end{align*}

Which of these conditions is satisfied depends on the coupling strength according to Eq. \eqref{eq:cond3}.  If $\kappa > |\beta - \alpha|/2$ then condition (i) must be satisfied, which is simply the traditional observation of strong spin-photon coupling.  However when $\kappa < |\beta - \alpha|/2$, condition (ii) is  satisfied and a resonance crossing is observed, even though $\kappa \ne 0$.  The transition from (i) to (ii) occurs when $\kappa = \kappa_\text{EP} = |\beta - \alpha|/2$ where both the resonance frequency and line width will merge.  This special value of the coupling strength defines the exceptional point, where the eigenvectors of the system coalesce [\!\!\citenum{Heiss1999, Heiss2000}].  This behaviour corresponds to a special topological structure defined by two intersecting Riemann sheets as shown in Fig. \ref{fig1} (a).  Here the coupling strength has been plotted in terms of the experimentally controlled variable, the angle $\theta$ between the local microwave and static magnetic fields, which will be discussed in detail below.  We see that for $\kappa > \kappa_\text{EP}$ (large $\theta$) the expected anti crossing is observed, as highlighted by the blue curve.  At the EP the two Riemann sheets cross and, as highlighted with the yellow curve, at $0 < \kappa < \kappa_\text{EP}$ (small $\theta$) a crossing is observed.  Fig. \ref{fig1} (b) shows similar behaviour in the line width, which above $\kappa_\text{EP}$ shows a crossing in blue while below $\kappa_\text{EP}$  shows an anti crossing in yellow.
\begin{figure}[t!]
\includegraphics[width = 8 cm]{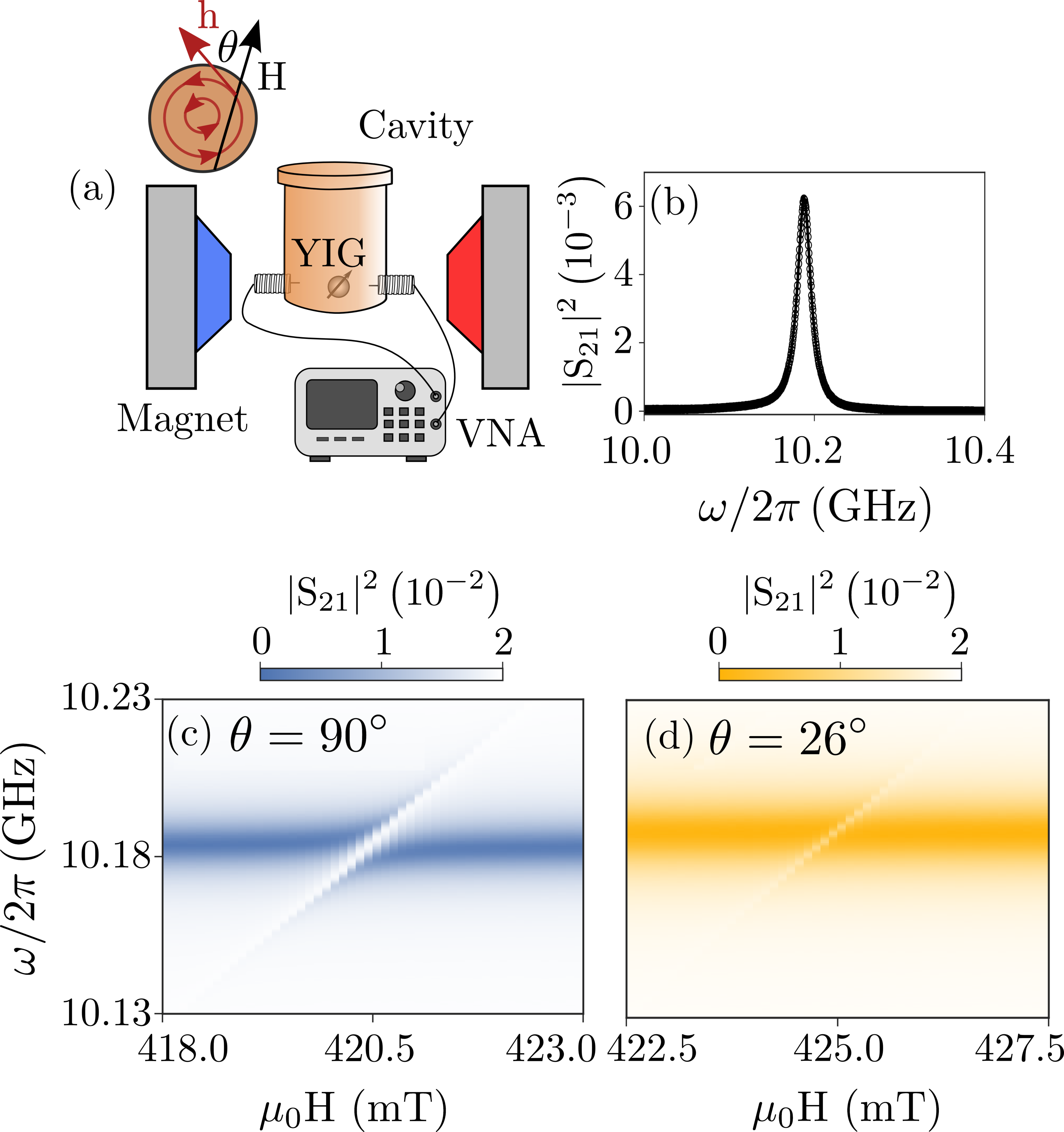}
\caption{(a) Experimental setup allowing in-situ control of the coupling strength by controlling the angle between the local microwave and static magnetic fields as illustrated in the inset.  (b) Transmission spectra of the TM$_{011}$ mode at $\mu_0$H = 0 mT.  Open symbols are experimental data and solid curve is a Lorentz fit.  (c) $\omega-H$ dispersion measurement at $\theta = 90^\circ > \theta_\text{EP}$ and (d) $\theta = 26^\circ < \theta_\text{EP}$.}
\label{fig2a}
\end{figure}

To experimentally observe the topological structure of the spin-photon system we placed a 0.3 mm YIG sphere at the outer, bottom edge of a cylindrical microwave cavity made of oxygen-free copper (diameter, height = 25, 33 mm) as sketched in Fig. \ref{fig2a} (a).  Two coaxial cables were connected to a vector network analyzer (VNA) and used to measure the complex microwave transmission spectra $S_{21}$.  The transmission amplitude, $|S_{21}|^{2}$, of the empty cavity is plotted in Fig. \ref{fig2a} (b) and shows a sharp peak at the TM$_{011}$ resonance frequency $\omega_c/2\pi = 10.183$ GHz.  This spectra is fit to a Lorentz function $S_{21} \propto \frac{\omega}{\omega - \omega_{c} + i\beta\omega_{c}}$ [\!\!\citenum{Harder2016b}], and a damping of $\beta = 7.6 \times 10^{-4}$ ($Q = 660$) is found.  An externally applied magnetic field is used to tune the FMR dispersion according to the Kittel equation, $\omega_{r} = \gamma(H + H_\text{A})$.  In our sample we find a YIG gyromagnetic ratio and anisotropy field of $\gamma$ = 28$\times 2\pi~\mu_{0}$GHz/T and $\mu_0H_\text{A} = -0.057$ T respectively and a Gilbert damping of $\alpha = 1.1 \times 10^{-4}$.  Based on this characterization of our spin-photon system, we can predict the presence of the EP at $\kappa_\text{EP} = 3.25 \times 10^{-4}$.  

In order to probe the full topology of the spin-photon eigenvalues we must tune the spin-photon coupling strength.  To do so we use the in-situ technique developed by Bai et al. [\!\!\citenum{Bai2016}], where the angle $\theta$ between the local microwave, $h$, and static, $H$, magnetic fields is controlled.  This angle changes the Zeeman interaction between the local microwave field and the spins and results in a coupling strength, $\kappa = \kappa_\text{M} |\sin\theta|$.  By measuring $\omega_\text{gap}$ at $\theta = 90^\circ$ we find that $\kappa_\text{M} = 5.9 \times 10^{-4}$ and therefore $\theta_\text{EP} = \arcsin(\kappa_\text{EP}/\kappa_\text{M}) = 33^\circ$.   

Fig. \ref{fig2a} (c) shows the transmission mapping at $\theta = 90^\circ > \theta_\text{EP}$, corresponding to the blue curve in the Riemann sheet of Fig. \ref{fig1} (a).  In this measurement strong coupling is achieved and the expected resonance anti crossing is observed.  On the other hand when we set $\theta = 26^\circ < \theta_\text{EP}$ the mapping shown in Fig. \ref{fig2a} (d), corresponding to the yellow curve in the Riemann sheet of Fig. \ref{fig1} (a), does not show an anti crossing and instead simply shows a dip at the FMR dispersion corresponding to traditional FMR absorption.  

\begin{figure}[tb]
 \includegraphics[width = 8cm]{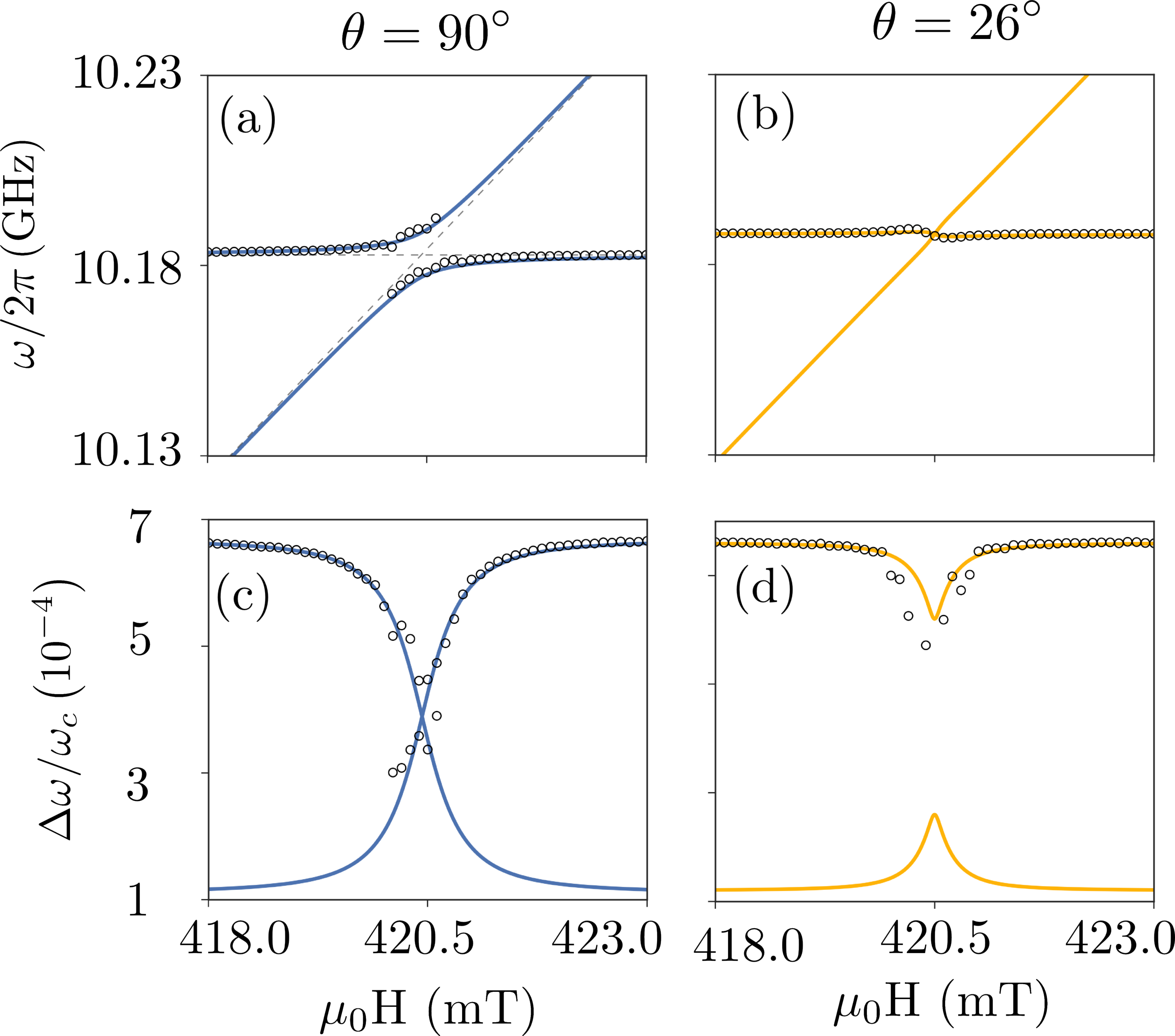}
\caption{(a) A resonance anti crossing and (c) a line width crossing is observed when $\theta = 90^\circ > \theta_\text{EP}$.  On the other hand, below the EP for $\theta = 26^\circ < \theta_\text{EP}$ we instead observe (b) a resonance crossing and (d) a line width anti crossing.  In all panels open circles are experimental data and solid curves are calculations according to Eq. \eqref{eq:eigen}.}
\label{fig2b}
\end{figure}

From Eq. \eqref{eq:hamiltonian} the microwave transmission can be found,
\begin{equation}
S_{21} \propto \frac{ (\omega-\omega_{r}+ i\alpha\omega_c)\omega_c}{(\omega - \omega_{c} + i\beta\omega_{c})(\omega - \omega_{r}+i\alpha\omega_c) - \kappa^{2}\omega_c^{2}}
\label{eq:s21}
\end{equation}
which, in the strong coupling regime, becomes Lorentzian near each resonance mode [\!\!\citenum{Huebl2013a, Harder2016b}].  By fitting each peak to a Lorentz function we can therefore determine the dispersion and line width for $\theta > \theta_\text{EP}$.  These fitting results are shown in Fig. \ref{fig2b} (a) (dispersion) and (c) (line width) as open symbols.  As expected we observe a dispersion anti crossing in Fig. \ref{fig2b} (a) and line width crossing in Fig. \ref{fig2b} (c).  The expected behaviour of the dispersion and line width can be calculated according to Eq. \eqref{eq:eigen} using the known cavity and FMR damping and resonance frequencies and a coupling strength estimated from the minimal frequency gap between eigenmodes [\!\!\citenum{Harder2016b}].  This calculation is shown as the solid blue curves in Fig. \ref{fig2b} (a) and (c) and agrees well with the experimental data.  

To extract the dispersion and line width information when $\theta < \theta_\text{EP}$ a similar analysis procedure can be used to fit the dominant peak.  For most fields only one Lorentz peak is observed and can be fit to determine the line width and resonance position.  However near the field where $\omega_c = \omega_r$ a peak splitting is observed corresponding to the FMR absorption, with a weak FMR-like peak behaving as an additional background.  Therefore when both peaks are present we fit the dominant peak using a linear combination of Lorentz and dispersive functions, accounting for the line shape distortion resulting from the antiresonance at $\omega_r$ [\!\!\citenum{Harder2016b, Harder2016}].  (Note that if no peak splitting was observed, as would happen at even lower coupling strengths, alternative approaches, such as a careful line width analysis, could be used to determine the coupling behaviour [\!\!\citenum{Philipp2000, Herskind2009}]).  The results of this fitting procedure are shown as open symbols in Fig. \ref{fig2b} (b) and (d).  Surprisingly in this case we observe a dispersion crossing and line width anti crossing.  By estimating the coupling strength according to $\kappa = \kappa_\text{M} |\sin\theta|$ we can again use Eq. \eqref{eq:eigen} to calculate the expected behaviour based on the topological structure of the eigenvalue spectrum.  These calculations are shown as the yellow solid curves in Fig. \ref{fig2b} (b) and (d) and agree well with the experimental results.

In light of this observation, that a resonance crossing may be observed even in the presence of non-zero coupling, it is important to examine the dispersion gap further.  Since $\Gamma_{1,2}$ characterize the losses of the spin-photon system, we must have $|\Gamma_1 - \Gamma_2| < |\alpha - \beta|$.  Therefore provided the coupling strength is large, that is $|\alpha - \beta| \ll \kappa$, the dispersion gap in Eq. \eqref{eq:cond3} will be proportional to the coupling strength, $\omega_\text{gap} = |\omega_1-\omega_2| \sim 2 \kappa \omega_c$.  This is the standard relation used to determine the coupling strength from the gap size.  However near the EP the situation changes and the coupling strength becomes comparable to the damping, $\kappa \sim |\alpha - \beta|$.  Therefore the dispersion gap can only be determined by the full expression of Eq. \eqref{eq:cond3}, however if we assume the presence of an anti crossing, i.e. $\Gamma_1 = \Gamma_2$, then $\omega_\text{gap} = \sqrt{4 \kappa^2 - \left(\beta - \alpha\right)^2} \omega_c$.  So even when an anti crossing is present, the coupling gap depends on the damping and this dependence becomes more important as the strong/weak transition is approached.  To estimate the magnitude of the cooperativity near the strong/weak transition we can evaluate $C$ at the EP where $\kappa = |\beta - \alpha|/2$ and therefore $C_\text{EP} = \kappa^2/\alpha\beta = 1/4\left(\alpha/\beta + \beta/\alpha\right)$.  As $C_\text{EP}$ is determined solely by the damping we see that if $\alpha/\beta \gtrsim 6$ (or $\beta/\alpha \gtrsim 6)$,  $C_\text{EP} > 1$, which means that a dispersion crossing would be observed although we are in the traditional strong coupling regime.  Indeed applying this estimation to our experiment we find that the resonance crossing observed in Fig. \ref{fig2b} (b) occurs at $C = 1.3$.  For this reason it may be more accurate to define the strong/weak transition using $\kappa > \kappa_\text{EP}$, rather than $C >1$.

\section{Geometric Mode Switching} \label{sec:phase}

\begin{figure}[t!]
 \includegraphics[width = 8 cm]{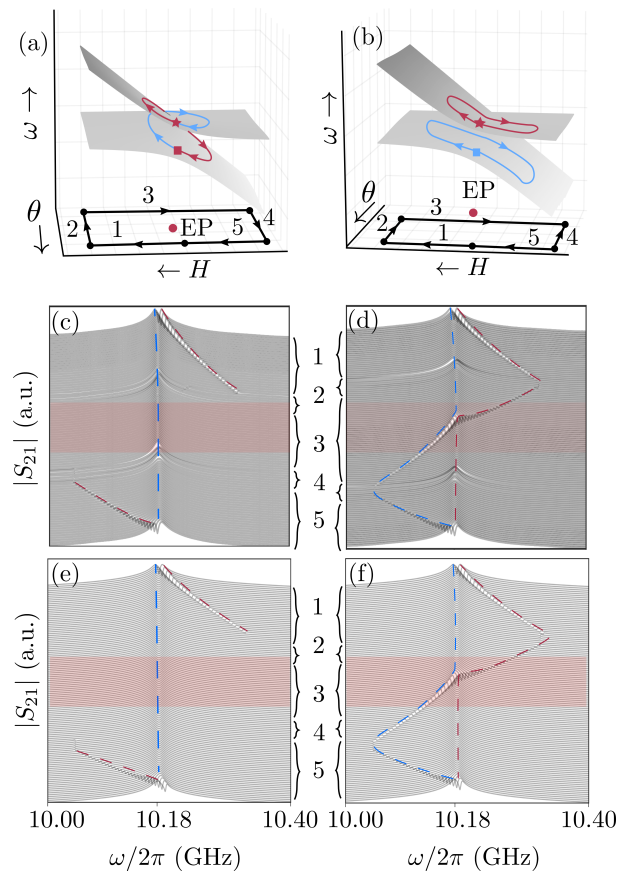}
\caption{(a) Path taken to encircle the EP.  A square (star) denotes the starting (ending) positions, while the red (blue) curve indicates the path taken by Mode 1 (Mode 2).  Note that the two paths are connected.  (b) Path taken without encircling the EP.  Each mode now behaves independently as shown with the red and blue curves.  (c) Experimental spectra observed while encircling the EP.  From top to bottom the system is tuned along paths $1 \to 5$.  One mode stays constant at the cavity frequency while the other mode initially shifts to high frequencies, eventually disappearing, until it reappears at lower frequencies.  Therefore the modes have switched positions.  (d) When the EP is not circled, an anti crossing occurs during Path 3 and therefore the modes maintain their relative orientation.  This key difference along Path 3 is highlighted with red shading.  (e) The theoretical spectra calculated according to Eq. \eqref{eq:s21} when the EP is encircled and (d) not encircled.}
\label{fig3}
\end{figure} 

The topological structure arising near an EP has interesting effects on the eigenvalues and eigenvectors when dynamically encircled.  Namely, upon encircling the EP the two modes will switch eigenvalues, while the eigenvectors will acquire a geometric phase.  These effects bear similarities to the phenomena of Berry's phase, however an important difference is that Berry's phase occurs in Hermitian systems and therefore the crossing of eigenvalues represents a true degeneracy in the spectrum where the two eigenvectors are still orthogonal.  On the other hand in a non-Hermitian system the eigenvectors are no longer required to be orthogonal and instead coalesce at the EP [\!\!\citenum{Heiss1999}].

To observe geometric mode switching in our spin-photon system we carefully tune the magnetic field and coupling strength in order to encircle the EP.  The path taken is shown in Fig. \ref{fig3} (a).  Starting from the coupling point (420.5 mT, $90^\circ$) we first increase the magnetic field along Path 1 to reach (430 mT, $90^\circ$).  The coupling strength is then decreased along Path 2 until we reach (430 mT, 15$^\circ$) which is beyond the EP.  We then decrease the magnetic field along Path 3 until (415 mT, 15$^\circ$), and increase the coupling strength along Path 4 until (415 mT, 90$^\circ$).  Finally we increase  the magnetic field along Path 5 back to the coupling point at (420.5 mT, $90^\circ$).  To follow the mode evolution along these paths we label the high frequency mode at the initial position (top spectra in Fig. \ref{fig3} (a)) Mode 1 and the low frequency mode Mode 2.  Note that the assignment of mode labels in this way is not meaningful when looking only at a single spectra, indeed each spectra will always show two modes and the frequency of these modes is path-independent and determined from Eq. \eqref{eq:eigen}.  However by carefully tuning the position in $\theta-H$ parameter space we can follow the evolution of the modes at any starting position and it is in this sense, to follow the evolution, that we assign such labels.  

In Fig. \ref{fig3} (a) we see that as the system is tuned along Path 1, by increasing the magnetic field, Mode 1 shifts out to high frequencies and the amplitude decreases.  Meanwhile Mode 2 remains fixed at the cavity frequency.  We then decrease the coupling strength along Path 2, decrease the field along Path 3 and increase the coupling strength along Path 4.  All this time Mode 2 remains at the cavity frequency while the Mode 1 amplitude is low and therefore difficult to see.  However as we increase the field along Path 5 back towards the starting point, we find that Mode 1 reappears at low frequencies while Mode 2 remains in the initial position and is therefore now at the higher frequency.  Therefore we observe mode switching due to encircling of the EP; the mode initially at low (high) frequencies has continuously evolved to appear at high (low) frequencies after a complete path around the EP.  The spectra along these paths can also be theoretically calculated according to Eq. \eqref{eq:s21} and are shown in Fig. \ref{fig3} (e).  Here we treat the proportionality constant, $\eta$, in Eq. \eqref{eq:s21} as a single fitting parameter, using $\eta = 1 \times 10^{-6}$ for all calculated spectra in Fig. \ref{fig3} (e).  The agreement between experiment and theory is excellent, with the small shifts in the experimental data along Paths 2 and 4 resulting from small frequency shifts in the cavity mode when the angle is tuned.

Such mode switching behaviour can be compared to a path where the EP is not encircled as shown in Fig. \ref{fig3} (b).  The path taken here is nearly the same as the one taken previously, however Path 3 is now at $\theta = 70^\circ$ so that $\theta > \theta_\text{EP}$.  The transmission spectra observed along this modified path are shown in Fig. \ref{fig3} (d) where top to bottom corresponds to $1 \to 5$.  In this case Mode 1 again shifts to high frequencies along Path 1 with Mode 2 fixed at $\omega_c$.  Along Path 2 and the first part of Path 3 Mode 2 again remains at the cavity mode frequency while Mode 1 shifts to low frequencies as the field is decreased.  However when the coupling point is encountered along Path 3, Mode 2 begins to shift towards low frequencies while Mode 1 remains fixed at $\omega_c$.  Along Path 4 and 5 Mode 1 remains at $\omega_c$ at a higher frequency than Mode 2.  Therefore in this case, when the EP is not encircled, Mode 2 remains at lower frequencies than Mode 1 for the entire path and no switching is observed.  Again we compare this experimental result to the calculated spectra along this path which is shown in Fig. \ref{fig3} (f) and find excellent agreement (again $\eta = 1 \times 10^{-6}$).
 
\section{Manipulating the Spin-Photon Topology} \label{sec:manipulate}

\begin{figure}[t!]
 \includegraphics[width = 8 cm]{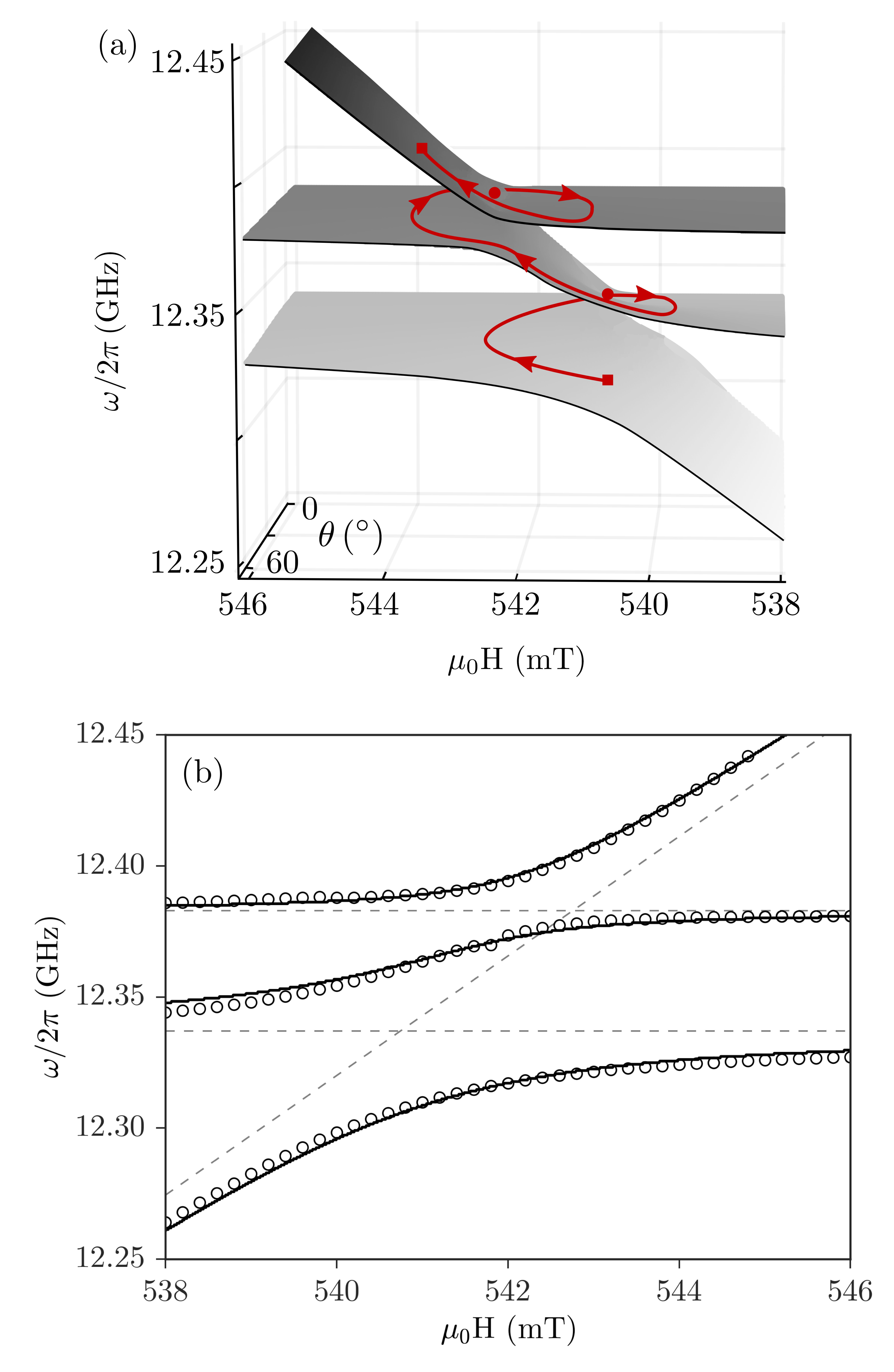}
\caption{(a) The topology of a three-mode spin-photon system with two cavity modes and one FMR mode.  The calculation is according to the eigenvalues of Eq. \ref{eq:3ModeHamiltonian} using our experimental parameters.  A path encircling both EPs, resembling an Archimedean spiral, is shown in red.  (b) Experimental observation of a single slice in the three-mode topology at $\theta = 61^\circ$.  Circles are experimental data and the black curve shows the calculated eigenvalues of Eq. \ref{eq:3ModeHamiltonian}.}
\label{fig4}
\end{figure}

An interesting possibility offered by the spin-photon system is to tune the topology with additional modes, for example using spin waves [\!\!\citenum{MaierFlaig2016, Zhang2016}], multiple FM materials [\!\!\citenum{Zhang2015g, Bai2017}], or additional cavity modes [\!\!\citenum{Hyde2016}].  The use of additional cavity modes is particularly intriguing as this can be realized in a height tuneable cavity which offers control over the mode frequencies and cavity damping which is not necessarily available with spin waves.  The Hamiltonian of such a three-level system consisting of one spin resonance and two cavity modes is a straight forward extension of the two-level system we have already considered in detail,
\begin{equation}
H = \left(\begin{array}{ccc}
\omega_{c1} - i \beta_1 \omega_{c1} & 0 & \kappa_1 \omega_{c1} \\
0 & \omega_{c2} - i \beta_2 \omega_{c1} & \kappa_2 \omega_{c1} \\
\kappa_1 \omega_{c1} & \kappa_2 \omega_{c1} & \omega_r - i \alpha \omega_{c1}
\end{array}\right). \label{eq:3ModeHamiltonian}
\end{equation}
The calculated eigenvalues of Eq. \ref{eq:3ModeHamiltonian} are plotted in Fig. \ref{fig4} (a), where two red circles indicate the EPs and the red curve indicates a path which encircles both EPs.  Such a path resembles an Archimedean spiral, and the height could be further increased by including additional cavity modes.  The fact that there are multiple EPs means that the geometric mode switching is also tuned and it is now necessary to encircle the EPs four times before returning to the same state.

To demonstrate the feasibility of this approach we use a height tuneable cylindrical cavity with diameter = 25 mm.  The height can be tuned between 24 and 45 mm and for our experiment is set at 36.5 mm where we observe a TM$_{012}$ mode at $\omega_{c1}/2\pi = 12.383$ GHz ($\beta_1 = 4.8 \times 10^{-4}$, $Q_1 = 1000$) and a TE$_{211}$ mode at $\omega_{c2}/2\pi = 12.337$ GHz ($\beta_2 = 2.7 \times 10^{-4}$, $Q_2 = 1900$).  These modes are coupled to a 1 mm YIG sphere which is measured to have $\alpha = 1 \times 10^{-4}, \gamma =28 \times 2\pi$ $\mu_0$GHz/T and $H_A = -0.08$ T in our configuration.

Fig \ref{fig4} (b) shows a slice of the three-mode topology at $\theta = 61^\circ$ measured using our height adjustable setup.  Here the open circles are experimental data while the solid curves are the calculated eigenvalues of the three mode Hamiltonian in Eq. \eqref{eq:3ModeHamiltonian}.  The good agreement between theory and experiment in the strong coupling regime suggests that further exploration near the strong/weak transition and the EPs holds interesting potential.  In particular, by fitting the data we find that the spin-photon coupling with TM$_{012}$ is approximately twice as large as with TE$_{211}$.  This means that the EP of the TM mode will occur at a smaller angle than the TE mode, which can already be observed in the calculation of Fig. \ref{fig4} (a).  Therefore in such a system, by carefully tuning the damping (e.g. by controlling the coaxial pin length) as well as the sample location to tune the coupling strength, it could be possible to observe mode switching at one EP, and mode crossing at another EP.  Recently it was demonstrated that the presence of an EP can produce an asymmetric switching effect, where certain modes injected into a waveguide will be preferentially transmitted depending on the direction of injection [\!\!\citenum{Doppler2016}].  The ability to add EPs and control the topology, as demonstrated in our spin-photon system, could provide additional functionality to such a switching scheme, by allowing control of the mode selectivity based on which EP is encircled.  We hope that this idea can be explored in detail in future works.

\section{Conclusions} \label{sec:conclusion}

In summary, we have examined the role of damping in a strongly-coupled spin-photon system.  We observed that even within the traditional strong coupling regime where the cooperativity exceeds one, a crossing of the dispersion may be observed.  Such a feature is controlled by the presence of an exceptional point in the eigenvalue spectrum, which is a general feature of non-Hermitian systems where the eigenmodes coalesce.  This observation, and its theoretical explanation, suggests that the boundary between the strong and weak coupling regimes should be defined based on the exceptional point location, rather than the cooperativity.  Furthermore we have demonstrated the geometric mode switching associated with encircling an exceptional point, resulting in an exchange of eigenvalues when returning to the same point in the resonance frequency-coupling parameter space.  Finally, the ability to easily tune the topology of the spin-photon system with additional modes demonstrates the potential of such systems for the exploration of non-Hermitian physics and novel switching protocols.  In particular it would be exciting to investigate the role of mode coalescence at the exceptional point on the spin current produced in a bilayer system.

\section*{Acknowlegements}
M. H. is partially supported by IODE Canada.  P. H. is supported by the UMGF program. This work was funded by NSERC, CFI, and NSFC (No. 11429401) grants (C.-M. H.). 

\bibliography{mainText.bbl}

\begin{thebibliography}{36}%
\makeatletter
\providecommand \@ifxundefined [1]{%
 \@ifx{#1\undefined}
}%
\providecommand \@ifnum [1]{%
 \ifnum #1\expandafter \@firstoftwo
 \else \expandafter \@secondoftwo
 \fi
}%
\providecommand \@ifx [1]{%
 \ifx #1\expandafter \@firstoftwo
 \else \expandafter \@secondoftwo
 \fi
}%
\providecommand \natexlab [1]{#1}%
\providecommand \enquote  [1]{``#1''}%
\providecommand \bibnamefont  [1]{#1}%
\providecommand \bibfnamefont [1]{#1}%
\providecommand \citenamefont [1]{#1}%
\providecommand \href@noop [0]{\@secondoftwo}%
\providecommand \href [0]{\begingroup \@sanitize@url \@href}%
\providecommand \@href[1]{\@@startlink{#1}\@@href}%
\providecommand \@@href[1]{\endgroup#1\@@endlink}%
\providecommand \@sanitize@url [0]{\catcode `\\12\catcode `\$12\catcode
  `\&12\catcode `\#12\catcode `\^12\catcode `\_12\catcode `\%12\relax}%
\providecommand \@@startlink[1]{}%
\providecommand \@@endlink[0]{}%
\providecommand \url  [0]{\begingroup\@sanitize@url \@url }%
\providecommand \@url [1]{\endgroup\@href {#1}{\urlprefix }}%
\providecommand \urlprefix  [0]{URL }%
\providecommand \Eprint [0]{\href }%
\providecommand \doibase [0]{http://dx.doi.org/}%
\providecommand \selectlanguage [0]{\@gobble}%
\providecommand \bibinfo  [0]{\@secondoftwo}%
\providecommand \bibfield  [0]{\@secondoftwo}%
\providecommand \translation [1]{[#1]}%
\providecommand \BibitemOpen [0]{}%
\providecommand \bibitemStop [0]{}%
\providecommand \bibitemNoStop [0]{.\EOS\space}%
\providecommand \EOS [0]{\spacefactor3000\relax}%
\providecommand \BibitemShut  [1]{\csname bibitem#1\endcsname}%
\let\auto@bib@innerbib\@empty
\bibitem [{\citenamefont {Soykal}\ and\ \citenamefont
  {Flatt{\'{e}}}(2010{\natexlab{a}})}]{Soykal2010}%
  \BibitemOpen
  \bibfield  {author} {\bibinfo {author} {\bibfnamefont {{\"{O}}.~O.}\
  \bibnamefont {Soykal}}\ and\ \bibinfo {author} {\bibfnamefont {M.~E.}\
  \bibnamefont {Flatt{\'{e}}}},\ }\href {\doibase 10.1103/PhysRevLett.104.077202} {\bibfield  {journal} {\bibinfo  {journal}
  {Phys. Rev. Lett.}\ }\textbf {\bibinfo {volume} {104}},\ \bibinfo {pages}
  {077202} (\bibinfo {year} {2010}{\natexlab{a}})},\ \Eprint
  {http://arxiv.org/abs/0907.3926} {arXiv:0907.3926} \BibitemShut {NoStop}%
\bibitem [{\citenamefont {Huebl}\ \emph {et~al.}(2013)\citenamefont {Huebl},
  \citenamefont {Zollitsch}, \citenamefont {Lotze}, \citenamefont {Hocke},
  \citenamefont {Greifenstein}, \citenamefont {Marx}, \citenamefont {Gross},\
  and\ \citenamefont {Goennenwein}}]{Huebl2013a}%
  \BibitemOpen
  \bibfield  {author} {\bibinfo {author} {\bibfnamefont {H.}~\bibnamefont
  {Huebl}}, \bibinfo {author} {\bibfnamefont {C.~W.}\ \bibnamefont
  {Zollitsch}}, \bibinfo {author} {\bibfnamefont {J.}~\bibnamefont {Lotze}},
  \bibinfo {author} {\bibfnamefont {F.}~\bibnamefont {Hocke}}, \bibinfo
  {author} {\bibfnamefont {M.}~\bibnamefont {Greifenstein}}, \bibinfo {author}
  {\bibfnamefont {A.}~\bibnamefont {Marx}}, \bibinfo {author} {\bibfnamefont
  {R.}~\bibnamefont {Gross}}, \ and\ \bibinfo {author} {\bibfnamefont
  {S.~T.~B.}\ \bibnamefont {Goennenwein}},\ }\href {\doibase 10.1103/PhysRevLett.111.127003} {\bibfield  {journal} {\bibinfo  {journal}
  {Phys. Rev. Lett.}\ }\textbf {\bibinfo {volume} {111}},\ \bibinfo {pages}
  {127003} (\bibinfo {year} {2013})},\ \Eprint {http://arxiv.org/abs/1207.6039}
  {arXiv:1207.6039} \BibitemShut {NoStop}%
\bibitem [{\citenamefont {Zhang}\ \emph {et~al.}(2014)\citenamefont {Zhang},
  \citenamefont {Zou}, \citenamefont {Jiang},\ and\ \citenamefont
  {Tang}}]{Zhang2014}%
  \BibitemOpen
  \bibfield  {author} {\bibinfo {author} {\bibfnamefont {X.}~\bibnamefont
  {Zhang}}, \bibinfo {author} {\bibfnamefont {C.-L.}\ \bibnamefont {Zou}},
  \bibinfo {author} {\bibfnamefont {L.}~\bibnamefont {Jiang}}, \ and\ \bibinfo
  {author} {\bibfnamefont {H.~X.}\ \bibnamefont {Tang}},\ }\href {\doibase 10.1103/PhysRevLett.113.156401} {\bibfield  {journal} {\bibinfo  {journal}
  {Phys. Rev. Lett.}\ }\textbf {\bibinfo {volume} {113}},\ \bibinfo {pages}
  {156401} (\bibinfo {year} {2014})},\ \Eprint {http://arxiv.org/abs/1405.7062}
  {arXiv:1405.7062} \BibitemShut {NoStop}%
\bibitem [{\citenamefont {Tabuchi}\ \emph {et~al.}(2014)\citenamefont
  {Tabuchi}, \citenamefont {Ishino}, \citenamefont {Ishikawa}, \citenamefont
  {Yamazaki}, \citenamefont {Usami},\ and\ \citenamefont
  {Nakamura}}]{Tabuchi2014a}%
  \BibitemOpen
  \bibfield  {author} {\bibinfo {author} {\bibfnamefont {Y.}~\bibnamefont
  {Tabuchi}}, \bibinfo {author} {\bibfnamefont {S.}~\bibnamefont {Ishino}},
  \bibinfo {author} {\bibfnamefont {T.}~\bibnamefont {Ishikawa}}, \bibinfo
  {author} {\bibfnamefont {R.}~\bibnamefont {Yamazaki}}, \bibinfo {author}
  {\bibfnamefont {K.}~\bibnamefont {Usami}}, \ and\ \bibinfo {author}
  {\bibfnamefont {Y.}~\bibnamefont {Nakamura}},\ }\href {\doibase 10.1103/PhysRevLett.113.083603} {\bibfield  {journal} {\bibinfo  {journal}
  {Phys. Rev. Lett.}\ }\textbf {\bibinfo {volume} {113}},\ \bibinfo {pages}
  {083603} (\bibinfo {year} {2014})},\ \Eprint {http://arxiv.org/abs/1405.1913}
  {arXiv:1405.1913} \BibitemShut {NoStop}%
\bibitem [{\citenamefont {Bai}\ \emph {et~al.}(2015)\citenamefont {Bai},
  \citenamefont {Harder}, \citenamefont {Chen}, \citenamefont {Fan},
  \citenamefont {Xiao},\ and\ \citenamefont {Hu}}]{Bai2015}%
  \BibitemOpen
  \bibfield  {author} {\bibinfo {author} {\bibfnamefont {L.}~\bibnamefont
  {Bai}}, \bibinfo {author} {\bibfnamefont {M.}~\bibnamefont {Harder}},
  \bibinfo {author} {\bibfnamefont {Y.~P.}\ \bibnamefont {Chen}}, \bibinfo
  {author} {\bibfnamefont {X.}~\bibnamefont {Fan}}, \bibinfo {author}
  {\bibfnamefont {J.~Q.}\ \bibnamefont {Xiao}}, \ and\ \bibinfo {author}
  {\bibfnamefont {C.-M.}\ \bibnamefont {Hu}},\ }\href {\doibase 10.1103/PhysRevLett.114.227201} {\bibfield  {journal} {\bibinfo  {journal}
  {Phys. Rev. Lett.}\ }\textbf {\bibinfo {volume} {114}},\ \bibinfo {pages}
  {227201} (\bibinfo {year} {2015})},\ \Eprint
  {http://arxiv.org/abs/1504.01335} {arXiv:1504.01335} \BibitemShut {NoStop}%
\bibitem [{\citenamefont {Zhang}\ \emph
  {et~al.}(2015{\natexlab{a}})\citenamefont {Zhang}, \citenamefont {Wang},
  \citenamefont {Li}, \citenamefont {Luo}, \citenamefont {Wu}, \citenamefont
  {Nori},\ and\ \citenamefont {You}}]{Zhang2015e}%
  \BibitemOpen
  \bibfield  {author} {\bibinfo {author} {\bibfnamefont {D.}~\bibnamefont
  {Zhang}}, \bibinfo {author} {\bibfnamefont {X.-M.}\ \bibnamefont {Wang}},
  \bibinfo {author} {\bibfnamefont {T.-F.}\ \bibnamefont {Li}}, \bibinfo
  {author} {\bibfnamefont {X.-Q.}\ \bibnamefont {Luo}}, \bibinfo {author}
  {\bibfnamefont {W.}~\bibnamefont {Wu}}, \bibinfo {author} {\bibfnamefont
  {F.}~\bibnamefont {Nori}}, \ and\ \bibinfo {author} {\bibfnamefont
  {J.}~\bibnamefont {You}},\ }\href {\doibase 10.1038/npjqi.2015.14} {\bibfield
   {journal} {\bibinfo  {journal} {npj Quantum Inf.}\ }\textbf {\bibinfo
  {volume} {1}},\ \bibinfo {pages} {15014} (\bibinfo {year}
  {2015}{\natexlab{a}})}\BibitemShut {NoStop}%
\bibitem [{\citenamefont {Haigh}\ \emph
  {et~al.}(2015{\natexlab{a}})\citenamefont {Haigh}, \citenamefont {Lambert},
  \citenamefont {Doherty},\ and\ \citenamefont {Ferguson}}]{Haigh2015a}%
  \BibitemOpen
  \bibfield  {author} {\bibinfo {author} {\bibfnamefont {J.~A.}\ \bibnamefont
  {Haigh}}, \bibinfo {author} {\bibfnamefont {N.~J.}\ \bibnamefont {Lambert}},
  \bibinfo {author} {\bibfnamefont {A.~C.}\ \bibnamefont {Doherty}}, \ and\
  \bibinfo {author} {\bibfnamefont {A.~J.}\ \bibnamefont {Ferguson}},\ }\href
  {\doibase 10.1103/PhysRevB.91.104410} {\bibfield  {journal} {\bibinfo
  {journal} {Phys. Rev. B}\ }\textbf {\bibinfo {volume} {91}},\ \bibinfo
  {pages} {104410} (\bibinfo {year} {2015}{\natexlab{a}})},\ \Eprint
  {http://arxiv.org/abs/1506.05631} {arXiv:1506.05631} \BibitemShut {NoStop}%
\bibitem [{\citenamefont {Cao}\ \emph {et~al.}(2015)\citenamefont {Cao},
  \citenamefont {Yan}, \citenamefont {Huebl}, \citenamefont {Goennenwein},\
  and\ \citenamefont {Bauer}}]{Cao2014}%
  \BibitemOpen
  \bibfield  {author} {\bibinfo {author} {\bibfnamefont {Y.}~\bibnamefont
  {Cao}}, \bibinfo {author} {\bibfnamefont {P.}~\bibnamefont {Yan}}, \bibinfo
  {author} {\bibfnamefont {H.}~\bibnamefont {Huebl}}, \bibinfo {author}
  {\bibfnamefont {S.~T.~B.}\ \bibnamefont {Goennenwein}}, \ and\ \bibinfo
  {author} {\bibfnamefont {G.~E.~W.}\ \bibnamefont {Bauer}},\ }\href {\doibase 10.1103/PhysRevB.91.094423} {\bibfield  {journal} {\bibinfo  {journal} {Phys.
  Rev. B}\ }\textbf {\bibinfo {volume} {91}},\ \bibinfo {pages} {094423}
  (\bibinfo {year} {2015})},\ \Eprint {http://arxiv.org/abs/1412.5809}
  {arXiv:1412.5809} \BibitemShut {NoStop}%
   \bibitem [{\citenamefont {Yao}\ \emph {et~al.}(2015)\citenamefont {Yao}}]{Yao2015}%
  \BibitemOpen
  \bibfield  {author} {\bibinfo {author} {\bibfnamefont {B. M.}~\bibnamefont
  {Yao}}, \bibinfo {author} {\bibfnamefont {Y. S.}~\bibnamefont
  {Gui}}, \bibinfo {author} {\bibfnamefont {Y.}~\bibnamefont
  {Xiao}}, \bibinfo {author} {\bibfnamefont {H.}~\bibnamefont
  {Guo}}, \bibinfo {author} {\bibfnamefont {X. S.}~\bibnamefont
  {Chen}}, \bibinfo {author} {\bibfnamefont {W.}~\bibnamefont
  {Lu}}, \bibinfo {author} {\bibfnamefont {C. L.}~\bibnamefont
  {Chien}}\ and\ \bibinfo {author} {\bibfnamefont {C.-M.}~\bibnamefont {Hu}},\
  }\href {\doibase 10.1103/physrevb.92.184407} {\bibfield  {journal} {\bibinfo
  {journal} {Phys. Rev. B}\ }\textbf {\bibinfo {volume} {92}},\ \bibinfo
  {pages} {184407} (\bibinfo {year} {2015})},\ \Eprint
  {http://arxiv.org/abs/1509.05804} {arXiv:1509.05804}\BibitemShut {NoStop}%
\bibitem [{\citenamefont {Bourhill}\ \emph
  {et~al.}(2016{\natexlab{a}})\citenamefont {Bourhill}, \citenamefont
  {Kostylev}, \citenamefont {Goryachev}, \citenamefont {Creedon},\ and\
  \citenamefont {Tobar}}]{Bourhill2015a}%
  \BibitemOpen
  \bibfield  {author} {\bibinfo {author} {\bibfnamefont {J.}~\bibnamefont
  {Bourhill}}, \bibinfo {author} {\bibfnamefont {N.}~\bibnamefont {Kostylev}},
  \bibinfo {author} {\bibfnamefont {M.}~\bibnamefont {Goryachev}}, \bibinfo
  {author} {\bibfnamefont {D.~L.}\ \bibnamefont {Creedon}}, \ and\ \bibinfo
  {author} {\bibfnamefont {M.~E.}\ \bibnamefont {Tobar}},\ }\href {\doibase 10.1103/PhysRevB.93.144420} {\bibfield  {journal} {\bibinfo  {journal} {Phys.
  Rev. B}\ }\textbf {\bibinfo {volume} {93}},\ \bibinfo {pages} {144420}
  (\bibinfo {year} {2016}{\natexlab{a}})},\ \Eprint
  {http://arxiv.org/abs/1512.07773v2} {arXiv:1512.07773} \BibitemShut
  {NoStop}%
\bibitem [{\citenamefont {Soykal}\ and\ \citenamefont
  {Flatt{\'{e}}}(2010{\natexlab{b}})}]{Soykal2010a}%
  \BibitemOpen
  \bibfield  {author} {\bibinfo {author} {\bibfnamefont {{\"{O}}.~O.}\
  \bibnamefont {Soykal}}\ and\ \bibinfo {author} {\bibfnamefont {M.~E.}\
  \bibnamefont {Flatt{\'{e}}}},\ }\href {\doibase 10.1103/PhysRevB.82.104413}
  {\bibfield  {journal} {\bibinfo  {journal} {Phys. Rev. B}\ }\textbf {\bibinfo
  {volume} {82}},\ \bibinfo {pages} {104413} (\bibinfo {year}
  {2010}{\natexlab{b}})},\ \Eprint {http://arxiv.org/abs/1005.3068}
  {arXiv:1005.3068} \BibitemShut {NoStop}%
\bibitem [{\citenamefont {Zhang}\ \emph
  {et~al.}(2015{\natexlab{b}})\citenamefont {Zhang}, \citenamefont {Zou},
  \citenamefont {Zhu}, \citenamefont {Marquardt}, \citenamefont {Jiang},\ and\
  \citenamefont {Tang}}]{Zhang2015g}%
  \BibitemOpen
  \bibfield  {author} {\bibinfo {author} {\bibfnamefont {X.}~\bibnamefont
  {Zhang}}, \bibinfo {author} {\bibfnamefont {C.-L.}\ \bibnamefont {Zou}},
  \bibinfo {author} {\bibfnamefont {N.}~\bibnamefont {Zhu}}, \bibinfo {author}
  {\bibfnamefont {F.}~\bibnamefont {Marquardt}}, \bibinfo {author}
  {\bibfnamefont {L.}~\bibnamefont {Jiang}}, \ and\ \bibinfo {author}
  {\bibfnamefont {H.~X.}\ \bibnamefont {Tang}},\ }\href {\doibase 10.1038/ncomms9914} {\bibfield  {journal} {\bibinfo  {journal} {Nat.
  Commun.}\ }\textbf {\bibinfo {volume} {6}},\ \bibinfo {pages} {8914}
  (\bibinfo {year} {2015}{\natexlab{b}})},\ \Eprint
  {http://arxiv.org/abs/1507.02791v1} {arXiv:1507.02791} \BibitemShut
  {NoStop}%
\bibitem [{\citenamefont {Tabuchi}\ \emph {et~al.}(2015)\citenamefont
  {Tabuchi}, \citenamefont {Ishino}, \citenamefont {Noguchi}, \citenamefont
  {Ishikawa}, \citenamefont {Yamazaki}, \citenamefont {Usami},\ and\
  \citenamefont {Nakamura}}]{Tabuchi2015b}%
  \BibitemOpen
  \bibfield  {author} {\bibinfo {author} {\bibfnamefont {Y.}~\bibnamefont
  {Tabuchi}}, \bibinfo {author} {\bibfnamefont {S.}~\bibnamefont {Ishino}},
  \bibinfo {author} {\bibfnamefont {A.}~\bibnamefont {Noguchi}}, \bibinfo
  {author} {\bibfnamefont {T.}~\bibnamefont {Ishikawa}}, \bibinfo {author}
  {\bibfnamefont {R.}~\bibnamefont {Yamazaki}}, \bibinfo {author}
  {\bibfnamefont {K.}~\bibnamefont {Usami}}, \ and\ \bibinfo {author}
  {\bibfnamefont {Y.}~\bibnamefont {Nakamura}},\ }\href {\doibase 10.1126/science.aaa3693} {\bibfield  {journal} {\bibinfo  {journal}
  {Science}\ }\textbf {\bibinfo {volume} {349}},\ \bibinfo {pages} {405}
  (\bibinfo {year} {2015})},\ \Eprint {http://arxiv.org/abs/1410.3781}
  {arXiv:1410.3781} \BibitemShut {NoStop}%
\bibitem [{\citenamefont {Hu}(2016)}]{Hu2015}%
  \BibitemOpen
  \bibfield  {author} {\bibinfo {author} {\bibfnamefont {C.-M.}\ \bibnamefont
  {Hu}},\ }\href@noop {} {\bibfield  {journal} {\bibinfo  {journal} {Phys. in
  Can.}\ }\textbf {\bibinfo {volume} {72}},\ \bibinfo {pages} {76} (\bibinfo
  {year} {2016})},\ \Eprint {http://arxiv.org/abs/1508.01966}
  {arXiv:1508.01966} \BibitemShut {NoStop}%
\bibitem [{\citenamefont {Bai}\ \emph {et~al.}(2017)\citenamefont {Bai},
  \citenamefont {Harder}, \citenamefont {Hyde}, \citenamefont {Zhang},
  \citenamefont {Hu}, \citenamefont {Chen},\ and\ \citenamefont
  {Xiao}}]{Bai2017}%
  \BibitemOpen
  \bibfield  {author} {\bibinfo {author} {\bibfnamefont {L.}~\bibnamefont
  {Bai}}, \bibinfo {author} {\bibfnamefont {M.}~\bibnamefont {Harder}},
  \bibinfo {author} {\bibfnamefont {P.}~\bibnamefont {Hyde}}, \bibinfo {author}
  {\bibfnamefont {Z.}~\bibnamefont {Zhang}}, \bibinfo {author} {\bibfnamefont
  {C.-M.}\ \bibnamefont {Hu}}, \bibinfo {author} {\bibfnamefont {Y.~P.}\
  \bibnamefont {Chen}}, \ and\ \bibinfo {author} {\bibfnamefont {J.~Q.}\
  \bibnamefont {Xiao}},\ }\href@noop {} {\bibfield  {journal} {\bibinfo
  {journal} {(unpublished)}\ }}\BibitemShut {NoStop}%
\bibitem [{\citenamefont {Haigh}\ \emph
  {et~al.}(2015{\natexlab{b}})\citenamefont {Haigh}, \citenamefont
  {Langenfeld}, \citenamefont {Lambert}, \citenamefont {Baumberg},
  \citenamefont {Ramsay}, \citenamefont {Nunnenkamp},\ and\ \citenamefont
  {Ferguson}}]{Haigh2015b}%
  \BibitemOpen
  \bibfield  {author} {\bibinfo {author} {\bibfnamefont {J.~A.}\ \bibnamefont
  {Haigh}}, \bibinfo {author} {\bibfnamefont {S.}~\bibnamefont {Langenfeld}},
  \bibinfo {author} {\bibfnamefont {N.~J.}\ \bibnamefont {Lambert}}, \bibinfo
  {author} {\bibfnamefont {J.~J.}\ \bibnamefont {Baumberg}}, \bibinfo {author}
  {\bibfnamefont {A.~J.}\ \bibnamefont {Ramsay}}, \bibinfo {author}
  {\bibfnamefont {A.}~\bibnamefont {Nunnenkamp}}, \ and\ \bibinfo {author}
  {\bibfnamefont {A.~J.}\ \bibnamefont {Ferguson}},\ }\href {\doibase 10.1103/PhysRevA.92.063845} {\bibfield  {journal} {\bibinfo  {journal} {Phys.
  Rev. A}\ }\textbf {\bibinfo {volume} {92}},\ \bibinfo {pages} {063845}
  (\bibinfo {year} {2015}{\natexlab{b}})},\ \Eprint
  {http://arxiv.org/abs/1510.06661} {arXiv:1510.06661} \BibitemShut {NoStop}%
\bibitem [{\citenamefont {Zhang}\ \emph
  {et~al.}(2016{\natexlab{a}})\citenamefont {Zhang}, \citenamefont {Zhu},
  \citenamefont {Zou},\ and\ \citenamefont {Tang}}]{Zhang2015b}%
  \BibitemOpen
  \bibfield  {author} {\bibinfo {author} {\bibfnamefont {X.}~\bibnamefont
  {Zhang}}, \bibinfo {author} {\bibfnamefont {N.}~\bibnamefont {Zhu}}, \bibinfo
  {author} {\bibfnamefont {C.-L.}\ \bibnamefont {Zou}}, \ and\ \bibinfo
  {author} {\bibfnamefont {H.~X.}\ \bibnamefont {Tang}},\ }\href {\doibase 10.1103/PhysRevLett.117.123605} {\bibfield  {journal} {\bibinfo  {journal}
  {Phys. Rev. Lett.}\ }\textbf {\bibinfo {volume} {117}},\ \bibinfo {pages}
  {123605} (\bibinfo {year} {2016}{\natexlab{a}})},\ \Eprint
  {http://arxiv.org/abs/1510.03545} {arXiv:1510.03545} \BibitemShut {NoStop}%
\bibitem [{\citenamefont {Osada}\ \emph {et~al.}(2016)\citenamefont {Osada},
  \citenamefont {Hisatomi}, \citenamefont {Noguchi}, \citenamefont {Tabuchi},
  \citenamefont {Yamazaki}, \citenamefont {Usami}, \citenamefont {Sadgrove},
  \citenamefont {Yalla}, \citenamefont {Nomura},\ and\ \citenamefont
  {Nakamura}}]{Osada2015}%
  \BibitemOpen
  \bibfield  {author} {\bibinfo {author} {\bibfnamefont {A.}~\bibnamefont
  {Osada}}, \bibinfo {author} {\bibfnamefont {R.}~\bibnamefont {Hisatomi}},
  \bibinfo {author} {\bibfnamefont {A.}~\bibnamefont {Noguchi}}, \bibinfo
  {author} {\bibfnamefont {Y.}~\bibnamefont {Tabuchi}}, \bibinfo {author}
  {\bibfnamefont {R.}~\bibnamefont {Yamazaki}}, \bibinfo {author}
  {\bibfnamefont {K.}~\bibnamefont {Usami}}, \bibinfo {author} {\bibfnamefont
  {M.}~\bibnamefont {Sadgrove}}, \bibinfo {author} {\bibfnamefont
  {R.}~\bibnamefont {Yalla}}, \bibinfo {author} {\bibfnamefont
  {M.}~\bibnamefont {Nomura}}, \ and\ \bibinfo {author} {\bibfnamefont
  {Y.}~\bibnamefont {Nakamura}},\ }\href {\doibase 10.1103/PhysRevLett.116.223601} {\bibfield  {journal} {\bibinfo  {journal}
  {Phys. Rev. Lett.}\ }\textbf {\bibinfo {volume} {116}},\ \bibinfo {pages}
  {223601} (\bibinfo {year} {2016})},\ \Eprint
  {http://arxiv.org/abs/1510.01837} {arXiv:1510.01837} \BibitemShut {NoStop}%
\bibitem [{\citenamefont {Bourhill}\ \emph
  {et~al.}(2016{\natexlab{b}})\citenamefont {Bourhill}, \citenamefont
  {Kostylev}, \citenamefont {Goryachev}, \citenamefont {Creedon},\ and\
  \citenamefont {Tobar}}]{Bourhill2015}%
  \BibitemOpen
  \bibfield  {author} {\bibinfo {author} {\bibfnamefont {J.}~\bibnamefont
  {Bourhill}}, \bibinfo {author} {\bibfnamefont {N.}~\bibnamefont {Kostylev}},
  \bibinfo {author} {\bibfnamefont {M.}~\bibnamefont {Goryachev}}, \bibinfo
  {author} {\bibfnamefont {D.~L.}\ \bibnamefont {Creedon}}, \ and\ \bibinfo
  {author} {\bibfnamefont {M.~E.}\ \bibnamefont {Tobar}},\ }\href {\doibase 10.1103/PhysRevB.93.144420} {\bibfield  {journal} {\bibinfo  {journal} {Phys.
  Rev. B}\ }\textbf {\bibinfo {volume} {93}},\ \bibinfo {pages} {144420}
  (\bibinfo {year} {2016}{\natexlab{b}})},\ \Eprint
  {http://arxiv.org/abs/1512.07773} {arXiv:1512.07773} \BibitemShut {NoStop}%
\bibitem [{\citenamefont {Liu}\ \emph {et~al.}(2016)\citenamefont {Liu},
  \citenamefont {Zhang}, \citenamefont {Tang},\ and\ \citenamefont
  {Flatt{\'{e}}}}]{Liu2016a}%
  \BibitemOpen
  \bibfield  {author} {\bibinfo {author} {\bibfnamefont {T.}~\bibnamefont
  {Liu}}, \bibinfo {author} {\bibfnamefont {X.}~\bibnamefont {Zhang}}, \bibinfo
  {author} {\bibfnamefont {H.~X.}\ \bibnamefont {Tang}}, \ and\ \bibinfo
  {author} {\bibfnamefont {M.~E.}\ \bibnamefont {Flatt{\'{e}}}},\ }\href
  {\doibase 10.1103/PhysRevB.94.060405} {\bibfield  {journal} {\bibinfo
  {journal} {Phys. Rev. B}\ }\textbf {\bibinfo {volume} {94}},\ \bibinfo
  {pages} {060405} (\bibinfo {year} {2016})},\ \Eprint
  {http://arxiv.org/abs/1604.07052} {arXiv:1604.07052} \BibitemShut {NoStop}%
\bibitem [{\citenamefont {Hisatomi}\ \emph {et~al.}(2016)\citenamefont
  {Hisatomi}, \citenamefont {Osada}, \citenamefont {Tabuchi}, \citenamefont
  {Ishikawa}, \citenamefont {Noguchi}, \citenamefont {Yamazaki}, \citenamefont
  {Usami},\ and\ \citenamefont {Nakamura}}]{Hisatomi2016}%
  \BibitemOpen
  \bibfield  {author} {\bibinfo {author} {\bibfnamefont {R.}~\bibnamefont
  {Hisatomi}}, \bibinfo {author} {\bibfnamefont {A.}~\bibnamefont {Osada}},
  \bibinfo {author} {\bibfnamefont {Y.}~\bibnamefont {Tabuchi}}, \bibinfo
  {author} {\bibfnamefont {T.}~\bibnamefont {Ishikawa}}, \bibinfo {author}
  {\bibfnamefont {A.}~\bibnamefont {Noguchi}}, \bibinfo {author} {\bibfnamefont
  {R.}~\bibnamefont {Yamazaki}}, \bibinfo {author} {\bibfnamefont
  {K.}~\bibnamefont {Usami}}, \ and\ \bibinfo {author} {\bibfnamefont
  {Y.}~\bibnamefont {Nakamura}},\ }\href {\doibase 10.1103/PhysRevB.93.174427}
  {\bibfield  {journal} {\bibinfo  {journal} {Phys. Rev. B}\ }\textbf {\bibinfo
  {volume} {93}},\ \bibinfo {pages} {174427} (\bibinfo {year} {2016})},\
  \Eprint {http://arxiv.org/abs/1601.03908} {arXiv:1601.03908} \BibitemShut
  {NoStop}%
\bibitem [{\citenamefont {Harder}\ \emph {et~al.}(2016)\citenamefont {Harder},
  \citenamefont {Bai}, \citenamefont {Match}, \citenamefont {Sirker},\ and\
  \citenamefont {Hu}}]{Harder2016b}%
  \BibitemOpen
  \bibfield  {author} {\bibinfo {author} {\bibfnamefont {M.}~\bibnamefont
  {Harder}}, \bibinfo {author} {\bibfnamefont {L.}~\bibnamefont {Bai}},
  \bibinfo {author} {\bibfnamefont {C.}~\bibnamefont {Match}}, \bibinfo
  {author} {\bibfnamefont {J.}~\bibnamefont {Sirker}}, \ and\ \bibinfo {author}
  {\bibfnamefont {C.}~\bibnamefont {Hu}},\ }\href {\doibase 10.1007/s11433-016-0228-6} {\bibfield  {journal} {\bibinfo  {journal} {Sci.
  China Physics, Mech. Astron.}\ }\textbf {\bibinfo {volume} {59}},\ \bibinfo
  {pages} {117511} (\bibinfo {year} {2016})},\ \Eprint
  {http://arxiv.org/abs/1601.06049v1} {arXiv:1601.06049} \BibitemShut
  {NoStop}%
\bibitem [{\citenamefont {Heiss}(1999)}]{Heiss1999}%
  \BibitemOpen
  \bibfield  {author} {\bibinfo {author} {\bibfnamefont {W.}~\bibnamefont
  {Heiss}},\ }\href {\doibase 10.1007/s100530050339} {\bibfield  {journal}
  {\bibinfo  {journal} {Eur. Phys. Jour. D}\ }\textbf {\bibinfo {volume} {7}},\
  \bibinfo {pages} {1} (\bibinfo {year} {1999})},\ \Eprint
  {http://arxiv.org/abs/quant-ph/9901023} {arXiv:quant-ph/9901023} \BibitemShut
  {NoStop}%
\bibitem [{\citenamefont {Heiss}(2000)}]{Heiss2000}%
  \BibitemOpen
  \bibfield  {author} {\bibinfo {author} {\bibfnamefont {W.~D.}\ \bibnamefont
  {Heiss}},\ }\href {\doibase 10.1103/PhysRevE.61.929} {\bibfield  {journal}
  {\bibinfo  {journal} {Phys. Rev. E}\ }\textbf {\bibinfo {volume} {61}},\
  \bibinfo {pages} {929} (\bibinfo {year} {2000})},\ \Eprint
  {http://arxiv.org/abs/quant-ph/9909047} {arXiv:quant-ph/9909047} \BibitemShut
  {NoStop}%
\bibitem [{\citenamefont {Bender}\ and\ \citenamefont
  {Boettcher}(1998)}]{Bender1998}%
  \BibitemOpen
  \bibfield  {author} {\bibinfo {author} {\bibfnamefont {C.~M.}\ \bibnamefont
  {Bender}}\ and\ \bibinfo {author} {\bibfnamefont {S.}~\bibnamefont
  {Boettcher}},\ }\href {\doibase 10.1103/PhysRevLett.80.5243} {\bibfield
  {journal} {\bibinfo  {journal} {Phys. Rev. Lett.}\ }\textbf {\bibinfo
  {volume} {80}},\ \bibinfo {pages} {5243} (\bibinfo {year} {1998})},\ \Eprint
  {http://arxiv.org/abs/physics/9712001} {arXiv:physics/9712001} \BibitemShut
  {NoStop}%
\bibitem [{\citenamefont {Bender}\ \emph {et~al.}(2007)\citenamefont {Bender},
  \citenamefont {Brody}, \citenamefont {Jones},\ and\ \citenamefont
  {Meister}}]{Bender2007}%
  \BibitemOpen
  \bibfield  {author} {\bibinfo {author} {\bibfnamefont {C.~M.}\ \bibnamefont
  {Bender}}, \bibinfo {author} {\bibfnamefont {D.~C.}\ \bibnamefont {Brody}},
  \bibinfo {author} {\bibfnamefont {H.~F.}\ \bibnamefont {Jones}}, \ and\
  \bibinfo {author} {\bibfnamefont {B.~K.}\ \bibnamefont {Meister}},\ }\href
  {\doibase 10.1103/PhysRevLett.98.040403} {\bibfield  {journal} {\bibinfo
  {journal} {Phys. Rev. Lett.}\ }\textbf {\bibinfo {volume} {98}},\ \bibinfo
  {pages} {040403} (\bibinfo {year} {2007})},\ \Eprint
  {http://arxiv.org/abs/quant-ph/0609032} {arXiv:quant-ph/0609032} \BibitemShut
  {NoStop}%
\bibitem [{\citenamefont {Philipp}\ \emph {et~al.}(2000)\citenamefont
  {Philipp}, \citenamefont {von Brentano}, \citenamefont {Pascovici},\ and\
  \citenamefont {Richter}}]{Philipp2000}%
  \BibitemOpen
  \bibfield  {author} {\bibinfo {author} {\bibfnamefont {M.}~\bibnamefont
  {Philipp}}, \bibinfo {author} {\bibfnamefont {P.}~\bibnamefont {von Brentano}},
  \bibinfo {author} {\bibfnamefont {G.}~\bibnamefont {Pascovici}}, \ and\
  \bibinfo {author} {\bibfnamefont {A.}~\bibnamefont {Richter}},\ }\href
  {\doibase 10.1103/PhysRevE.62.1922} {\bibfield  {journal} {\bibinfo
  {journal} {Phys. Rev. E}\ }\textbf {\bibinfo {volume} {62}},\ \bibinfo
  {pages} {1922} (\bibinfo {year} {2000})}\BibitemShut {NoStop}%
\bibitem [{\citenamefont {Dembowski}\ \emph {et~al.}(2001)\citenamefont
  {Dembowski}, \citenamefont {Gr\"af}, \citenamefont {Harney}, \citenamefont
  {Heine}, \citenamefont {Heiss}, \citenamefont {Rehfeld},\ and\ \citenamefont
  {Richter}}]{Dembowski2001}%
  \BibitemOpen
  \bibfield  {author} {\bibinfo {author} {\bibfnamefont {C.}~\bibnamefont
  {Dembowski}}, \bibinfo {author} {\bibfnamefont {H.~D.}\ \bibnamefont
  {Gr\"af}}, \bibinfo {author} {\bibfnamefont {H.~L.}\ \bibnamefont {Harney}},
  \bibinfo {author} {\bibfnamefont {A.}~\bibnamefont {Heine}}, \bibinfo
  {author} {\bibfnamefont {W.~D.}\ \bibnamefont {Heiss}}, \bibinfo {author}
  {\bibfnamefont {H.}~\bibnamefont {Rehfeld}}, \ and\ \bibinfo {author}
  {\bibfnamefont {A.}~\bibnamefont {Richter}},\ }\href {\doibase 10.1103/PhysRevLett.86.787} {\bibfield  {journal} {\bibinfo  {journal} {Phys.
  Rev. Lett.}\ }\textbf {\bibinfo {volume} {86}},\ \bibinfo {pages} {787}
  (\bibinfo {year} {2001})} \BibitemShut {NoStop}%
\bibitem [{\citenamefont {Liertzer}\ \emph {et~al.}(2012)\citenamefont
  {Liertzer}, \citenamefont {Ge}, \citenamefont {Cerjan}, \citenamefont
  {Stone}, \citenamefont {T{\"{u}}reci},\ and\ \citenamefont
  {Rotter}}]{Liertzer2012}%
  \BibitemOpen
  \bibfield  {author} {\bibinfo {author} {\bibfnamefont {M.}~\bibnamefont
  {Liertzer}}, \bibinfo {author} {\bibfnamefont {L.}~\bibnamefont {Ge}},
  \bibinfo {author} {\bibfnamefont {A.}~\bibnamefont {Cerjan}}, \bibinfo
  {author} {\bibfnamefont {A.~D.}\ \bibnamefont {Stone}}, \bibinfo {author}
  {\bibfnamefont {H.~E.}\ \bibnamefont {T{\"{u}}reci}}, \ and\ \bibinfo
  {author} {\bibfnamefont {S.}~\bibnamefont {Rotter}},\ }\href {\doibase 10.1103/PhysRevLett.108.173901} {\bibfield  {journal} {\bibinfo  {journal}
  {Phys. Rev. Lett.}\ }\textbf {\bibinfo {volume} {108}},\ \bibinfo {pages}
  {173901} (\bibinfo {year} {2012})},\ \Eprint {http://arxiv.org/abs/1109.0454}
  {arXiv:1109.0454} \BibitemShut {NoStop}%
\bibitem [{\citenamefont {Xu}\ \emph {et~al.}(2016)\citenamefont {Xu},
  \citenamefont {Mason}, \citenamefont {Jiang},\ and\ \citenamefont
  {Harris}}]{Xu2016}%
  \BibitemOpen
  \bibfield  {author} {\bibinfo {author} {\bibfnamefont {H.}~\bibnamefont
  {Xu}}, \bibinfo {author} {\bibfnamefont {D.}~\bibnamefont {Mason}}, \bibinfo
  {author} {\bibfnamefont {L.}~\bibnamefont {Jiang}}, \ and\ \bibinfo {author}
  {\bibfnamefont {J.~G.~E.}\ \bibnamefont {Harris}},\ }\href {\doibase 10.1038/nature18604} {\bibfield  {journal} {\bibinfo  {journal} {Nature}\
  }\textbf {\bibinfo {volume} {537}},\ \bibinfo {pages} {80} (\bibinfo {year}
  {2016})},\ \Eprint {http://arxiv.org/abs/1602.06881} {arXiv:1602.06881}
  \BibitemShut {NoStop}%
\bibitem [{\citenamefont {Graefe}\ \emph {et~al.}(2013)\citenamefont {Graefe},
  \citenamefont {Mailybaev},\ and\ \citenamefont {Moiseyev}}]{Graefe2013}%
  \BibitemOpen
  \bibfield  {author} {\bibinfo {author} {\bibfnamefont {E.-M.}\ \bibnamefont
  {Graefe}}, \bibinfo {author} {\bibfnamefont {A.~A.}\ \bibnamefont
  {Mailybaev}}, \ and\ \bibinfo {author} {\bibfnamefont {N.}~\bibnamefont
  {Moiseyev}},\ }\href {\doibase 10.1103/PhysRevA.88.033842} {\bibfield
  {journal} {\bibinfo  {journal} {Phys. Rev. A}\ }\textbf {\bibinfo {volume}
  {88}},\ \bibinfo {pages} {033842} (\bibinfo {year} {2013})},\ \Eprint
  {http://arxiv.org/abs/1207.5235} {arXiv:1207.5235} \BibitemShut {NoStop}%
\bibitem [{\citenamefont {Doppler}\ \emph {et~al.}(2016)\citenamefont
  {Doppler}, \citenamefont {Mailybaev}, \citenamefont {B{\"{o}}hm},
  \citenamefont {Kuhl}, \citenamefont {Girschik}, \citenamefont {Libisch},
  \citenamefont {Milburn}, \citenamefont {Rabl}, \citenamefont {Moiseyev},\
  and\ \citenamefont {Rotter}}]{Doppler2016}%
  \BibitemOpen
  \bibfield  {author} {\bibinfo {author} {\bibfnamefont {J.}~\bibnamefont
  {Doppler}}, \bibinfo {author} {\bibfnamefont {A.~A.}\ \bibnamefont
  {Mailybaev}}, \bibinfo {author} {\bibfnamefont {J.}~\bibnamefont
  {B{\"{o}}hm}}, \bibinfo {author} {\bibfnamefont {U.}~\bibnamefont {Kuhl}},
  \bibinfo {author} {\bibfnamefont {A.}~\bibnamefont {Girschik}}, \bibinfo
  {author} {\bibfnamefont {F.}~\bibnamefont {Libisch}}, \bibinfo {author}
  {\bibfnamefont {T.~J.}\ \bibnamefont {Milburn}}, \bibinfo {author}
  {\bibfnamefont {P.}~\bibnamefont {Rabl}}, \bibinfo {author} {\bibfnamefont
  {N.}~\bibnamefont {Moiseyev}}, \ and\ \bibinfo {author} {\bibfnamefont
  {S.}~\bibnamefont {Rotter}},\ }\href {\doibase 10.1038/nature18605}
  {\bibfield  {journal} {\bibinfo  {journal} {Nature}\ }\textbf {\bibinfo
  {volume} {537}},\ \bibinfo {pages} {76} (\bibinfo {year} {2016})},\ \Eprint
  {http://arxiv.org/abs/1603.02325} {arXiv:1603.02325} \BibitemShut {NoStop}%
\bibitem [{\citenamefont {Bai}\ \emph {et~al.}(2016)\citenamefont
  {Bai}, \citenamefont {Blanchette}, \citenamefont {Harder},
  \citenamefont {Chen}, \citenamefont {Fan}, \citenamefont {Xiao},\
  and\ \citenamefont {Hu}}]{Bai2016}%
\BibitemOpen
  \bibfield  {author} {\bibinfo {author} {\bibfnamefont {L.}~\bibnamefont
  {Bai}}, \bibinfo {author} {\bibfnamefont {K.}~\bibnamefont {Blanchette}},
  \bibinfo {author} {\bibfnamefont {M.}~\bibnamefont {Harder}}, \bibinfo
  {author} {\bibfnamefont {Y.~P.}\ \bibnamefont {Chen}}, \bibinfo {author}
  {\bibfnamefont {X.}~\bibnamefont {Fan}}, \bibinfo {author} {\bibfnamefont
  {J.~Q.}\ \bibnamefont {Xiao}}, \ and\ \bibinfo {author} {\bibfnamefont
  {C.~M.}\ \bibnamefont {Hu}},\ }\href {\doibase 10.1109/TMAG.2016.2527691}
  {\bibfield  {journal} {\bibinfo  {journal} {IEEE Trans. Magn.}\ }\textbf
  {\bibinfo {volume} {52}},\ \bibinfo {pages} {1000107} (\bibinfo {year}
  {2016})}\BibitemShut {NoStop}%
\bibitem [{\citenamefont {Herskind}\ \emph {et~al.}(2009)\citenamefont
  {Herskind}, \citenamefont {Dantan}, \citenamefont {Marler},
  \citenamefont {Albert},\
  and\ \citenamefont {Drewsen}}]{Herskind2009}%
  \BibitemOpen
  \bibfield  {author} {\bibinfo {author} {\bibfnamefont {P.}~\bibnamefont
  {Herskind}}, \bibinfo {author} {\bibfnamefont {A.}~\bibnamefont {Dantan}},
  \bibinfo {author} {\bibfnamefont {J.}~\bibnamefont {Marler}}, \bibinfo
  {author} {\bibfnamefont {M.}\ \bibnamefont {Albert}}, \ and\ \bibinfo {author} {\bibfnamefont
  {M.}\ \bibnamefont {Drewsen}},\ }\href {\doibase 10.1038/nphys1302}
  {\bibfield  {journal} {\bibinfo  {journal} {Nature Physics}\ }\textbf
  {\bibinfo {volume} {5}},\ \bibinfo {pages} {494} (\bibinfo {year}
  {2009})}\BibitemShut {NoStop}%
  \bibitem [{\citenamefont {Harder}\ \emph {et~al.}(2016)\citenamefont
  {Harder}, \citenamefont {Hyde}, \citenamefont {Bai}, \citenamefont
  {Match},\ and\ \citenamefont {Hu}}]{Harder2016}%
  \BibitemOpen
  \bibfield  {author} {\bibinfo {author} {\bibfnamefont {M.}~\bibnamefont
  {Harder}}, \bibinfo {author} {\bibfnamefont {P.}~\bibnamefont {Hyde}},
  \bibinfo {author} {\bibfnamefont {L.}~\bibnamefont {Bai}}, \bibinfo
  {author} {\bibfnamefont {C.}~\bibnamefont {Match}}, \ and\ \bibinfo {author}
  {\bibfnamefont {C.-M.}\ \bibnamefont {Hu}},\ }\href {\doibase 10.1103/PhysRevB.94.054403} {\bibfield  {journal} {\bibinfo  {journal} {Phys.
  Rev. B}\ }\textbf {\bibinfo {volume} {94}},\ \bibinfo {pages} {054403}
  (\bibinfo {year} {2016})},\ \Eprint {http://arxiv.org/abs/1606.03056}
  {arXiv:1606.03056} \BibitemShut {NoStop}%
\bibitem [{\citenamefont {Maier-Flaig}\ \emph {et~al.}(2016)\citenamefont
  {Maier-Flaig}, \citenamefont {Harder}, \citenamefont {Gross}, \citenamefont
  {Huebl},\ and\ \citenamefont {Goennenwein}}]{MaierFlaig2016}%
  \BibitemOpen
  \bibfield  {author} {\bibinfo {author} {\bibfnamefont {H.}~\bibnamefont
  {Maier-Flaig}}, \bibinfo {author} {\bibfnamefont {M.}~\bibnamefont {Harder}},
  \bibinfo {author} {\bibfnamefont {R.}~\bibnamefont {Gross}}, \bibinfo
  {author} {\bibfnamefont {H.}~\bibnamefont {Huebl}}, \ and\ \bibinfo {author}
  {\bibfnamefont {S.~T.~B.}\ \bibnamefont {Goennenwein}},\ }\href {\doibase 10.1103/PhysRevB.94.054433} {\bibfield  {journal} {\bibinfo  {journal} {Phys.
  Rev. B}\ }\textbf {\bibinfo {volume} {94}},\ \bibinfo {pages} {054433}
  (\bibinfo {year} {2016})},\ \Eprint {http://arxiv.org/abs/1601.05681}
  {arXiv:1601.05681} \BibitemShut {NoStop}%
\bibitem [{\citenamefont {Zhang}\ \emph
  {et~al.}(2016{\natexlab{b}})\citenamefont {Zhang}, \citenamefont {Zou},
  \citenamefont {Jiang},\ and\ \citenamefont {Tang}}]{Zhang2016}%
  \BibitemOpen
  \bibfield  {author} {\bibinfo {author} {\bibfnamefont {X.}~\bibnamefont
  {Zhang}}, \bibinfo {author} {\bibfnamefont {C.}~\bibnamefont {Zou}}, \bibinfo
  {author} {\bibfnamefont {L.}~\bibnamefont {Jiang}}, \ and\ \bibinfo {author}
  {\bibfnamefont {H.~X.}\ \bibnamefont {Tang}},\ }\href {\doibase 10.1063/1.4939134} {\bibfield  {journal} {\bibinfo  {journal} {J. Appl.
  Phys.}\ }\textbf {\bibinfo {volume} {119}},\ \bibinfo {pages} {023905}
  (\bibinfo {year} {2016}{\natexlab{b}})}\BibitemShut {NoStop}%
\bibitem [{\citenamefont {Hyde}\ \emph {et~al.}(2016)\citenamefont {Hyde},
  \citenamefont {Bai}, \citenamefont {Harder}, \citenamefont {Match},\ and\
  \citenamefont {Hu}}]{Hyde2016}%
  \BibitemOpen
  \bibfield  {author} {\bibinfo {author} {\bibfnamefont {P.}~\bibnamefont
  {Hyde}}, \bibinfo {author} {\bibfnamefont {L.}~\bibnamefont {Bai}}, \bibinfo
  {author} {\bibfnamefont {M.}~\bibnamefont {Harder}}, \bibinfo {author}
  {\bibfnamefont {C.}~\bibnamefont {Match}}, \ and\ \bibinfo {author}
  {\bibfnamefont {C.-M.}\ \bibnamefont {Hu}},\ }\href {\doibase 10.1063/1.4964602} {\bibfield  {journal} {\bibinfo  {journal} {Appl. Phys.
  Lett.}\ }\textbf {\bibinfo {volume} {109}},\ \bibinfo {pages} {152405}
  (\bibinfo {year} {2016})},\ \Eprint {http://arxiv.org/abs/1606.03469}
  {arXiv:1606.03469} \BibitemShut {NoStop}%
\end{thebibliography}%

\end{document}